\documentclass[sigconf]{acmart} 

\AtBeginDocument{%
  \providecommand\BibTeX{{%
    \normalfont B\kern-0.5em{\scshape i\kern-0.25em b}\kern-0.8em\TeX}}}

\setcopyright{none}
\copyrightyear{2023}
\acmYear{2023}
\acmDOI{10.1145/3544549.3574165}

\acmConference[CHI '23]{CHI Conference on Human Factors in Computing Systems Extended Abstracts}{April 29-May 5, 2023}{Hamburg, Germany}

\acmBooktitle{CHI Conference on Human Factors in Computing Systems (CHI '23), April 23-28, 2023, Hamburg, Germany}

\acmPrice{15.00}
\acmISBN{978-1-4503-XXXX-X/23/04}

\acmSubmissionID{C-XYZ}

\copyrightyear{2023}
\acmYear{2023}
\setcopyright{rightsretained}
\acmConference[CHI EA '23]{Extended Abstracts of the 2023 CHI Conference on Human Factors in Computing Systems}{April 23--28, 2023}{Hamburg, Germany}
\acmBooktitle{Extended Abstracts of the 2023 CHI Conference on Human Factors in Computing Systems (CHI EA '23), April 23--28, 2023, Hamburg, Germany}\acmDOI{10.1145/3544549.3574165}
\acmISBN{978-1-4503-9422-2/23/04}

\begin{document}

\title{Empirical Research Methods for Human-Computer Interaction}

\author{I. Scott MacKenzie}
\email{mack@yorku.ca}
\affiliation{%
  \institution{York University}
  \streetaddress{4700 Keele Street}
  \city{Toronto}
  \state{Ontario}
  \country{Canada}
  \postcode{M3J 1P3}
}

\author{Janet R. Read}
 \email{JCRead@uclan.ac.uk}
 \affiliation{%
  \institution{University of Central Lancashire}
  \streetaddress{xxx}
  \city{Preston}
  \country{UK} 
}

\author{Matthew Horton}
\email{MPLHorton@uclan.ac.uk}
\affiliation{%
  \institution{University of Central Lancashire}
  \city{Preston}
  \country{UK}
}


\begin{abstract}
 Most attendees at CHI conferences will agree that an experiment (user study) is the hallmark of good research in human-computer interaction. But what constitutes an experiment? And how does one go from an experiment to a CHI paper?

This course will teach how to pose testable research questions, how to make and measure observations, and how to design and conduct an experiment. Specifically, attendees will participate in a real experiment to gain experience as both an investigator and as a participant. The second session covers the statistical tools typically used to analyze data. Most notably, attendees will learn how to organize experiment results and write a CHI paper. 
\end{abstract}



\begin{teaserfigure}
  \includegraphics[width=\textwidth]{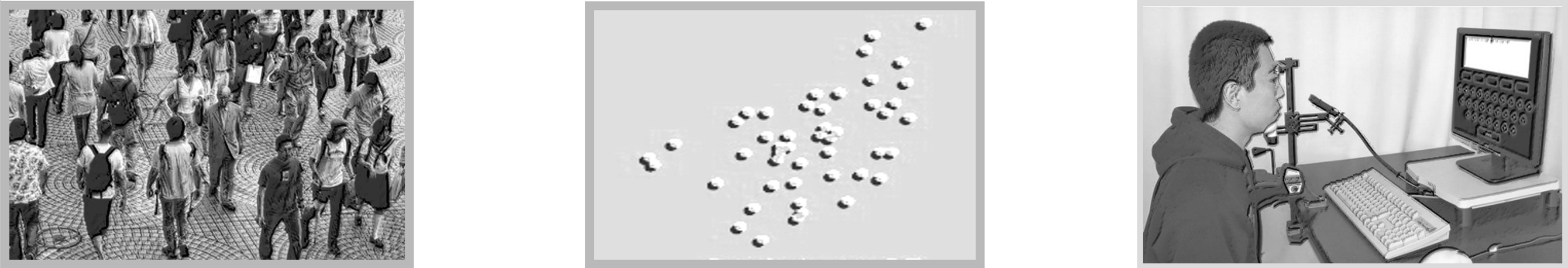}
  \caption{Empirical Research Methods. From left, Observational, Correlational, Experimental.}
  \Description{Empirical Research Methods. From left, Observational, Correlational, Experimental.  The figure is in three parts with the part on the left showing lots of people in a crowd.  This represents observational research.  The part in the middle shows some data points in a plot.  This represents correlational research.  The part on the right shows a subject in front of a computer with an eye tracking apparatus.  This represents experimental research.}
  \label{fig:teaser}
\end{teaserfigure}

\maketitle

\section{Benefits}

In this two-session course, attendees will learn how to conduct empirical research in human-computer interaction (HCI). This course delivers an A-to-Z tutorial on designing and doing a user study and demonstrates how to write a successful CHI paper. It would benefit anyone interested in conducting a user study or writing a CHI paper. Only a general HCI knowledge is required. 

\section{Intended Audience(s)}

 This course caters to attendees who are motivated to learn about, and use, empirical research methods in HCI research. Specifically, it is for those in academia or industry who evaluate interaction techniques using quantitative methods, or those who make decisions based on usability tests, and, in particular, user studies following an experimental methodology.

Approximately 75 attendees is the maximum practical size for this course. If the number of registrations is large, the instructors may consider teaching the course multiple times. 

\section{Prerequisites} 

No specific background is required other than a general knowledge of human-computer interaction as conveyed, for example, through an undergraduate HCI course or attendance at CHI conferences. Knowing how to enter formulae in a Microsoft Excel spreadsheet to compute means, standard deviations, etc., would be an asset. Knowledge of advanced statistics, such as the analysis of variance, is NOT required. Additionally, there is no linkage between this and any other CHI course. 

\section{Course History}

This course was offered at CHI 2007 (San Jose), CHI 2008 (Florence), CHI 2009 (Boston), CHI 2010 (Atlanta), CHI 2011 (Vancouver), CHI 2012 (Austin), CHI 2013 (Paris), CHI 2014 (Toronto), CHI 2016 (San Jose), CHI 2017 (Denver), CHI 2018 (Montreal), and CHI 2019 (Glasgow).  In addition, extended versions of this course have been given at the University of Tampere (Finland), the University of Central Lancashire (UK),  the University of Oslo (Norway), ETH Z\"urch (Switzerland), the University of the Balearic Islands (Spain), the IT University (Copenhagen, Denmark), Technical University of Denmark (Lyngby, Denmark), and the University of Aalborg (Denmark).\footnote{Please contact Scott MacKenzie, mack@yorku,ca, to discuss possibilities for your lab or institute.}  

\section{Content}

This course presents selected topics from Chapter 4 (Scientific Foundations), Chapter 5 (Designing HCI Experiments), and Chapter 6 (Hypothesis Testing) in \textit{Human-Computer Interaction: An Empirical Research Perspective} \cite{mackenzie2013a}.

Session 1 topics:

\begin{itemize}
\item What is empirical research and what is the scientific method (see Fig.~\ref{fig:teaser})?

\item Formulating "testable" research questions

\item How to design an experiment (broadly speaking) to answer research questions

\item Parts of an experiment (independent variables, dependent variables, counterbalancing, ethics approval, etc.)

\item Group participation in a real experiment \end{itemize}

Session 2 topics:

\begin{itemize}

\item Results and discussion of the experiment from session 1 (this affords a strong opportunity to revisit and expand on the elements of empirical research)

\item Experiment design issues ("within subjects" vs. "between subjects" factors, internal validity, external validity, counterbalancing test conditions, etc.)

\item Data analyses (main effects and interaction effects, requirements to establish cause and effect relationships, etc.)

\item How to organize and write a successful CHI paper (including suggestions for style and approach, as per CHI conference submissions) 
    
\end{itemize}

\section{Practical work}
\begin{figure}
\includegraphics[width=0.47\textwidth]{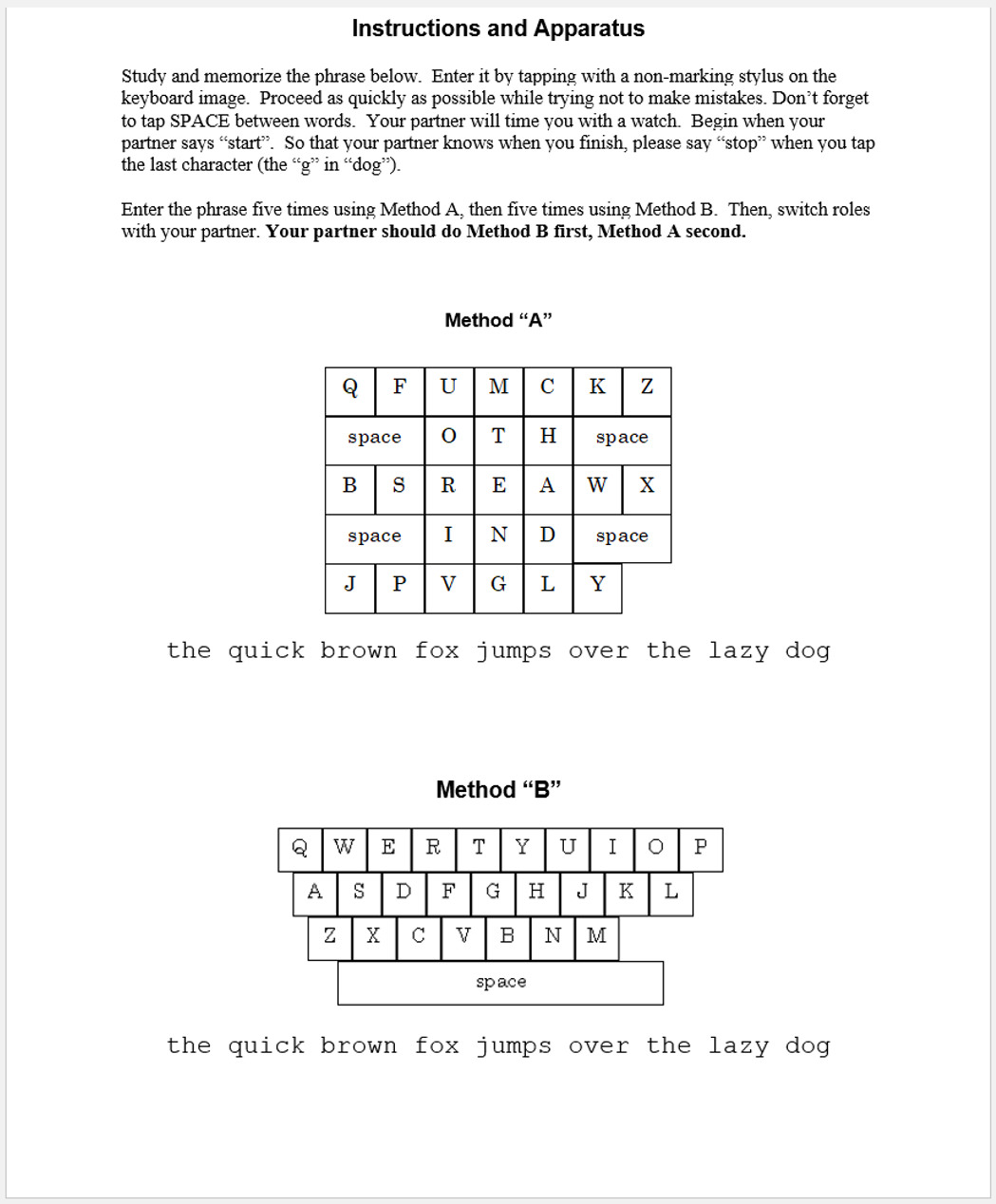}
\includegraphics[width=0.47\textwidth]{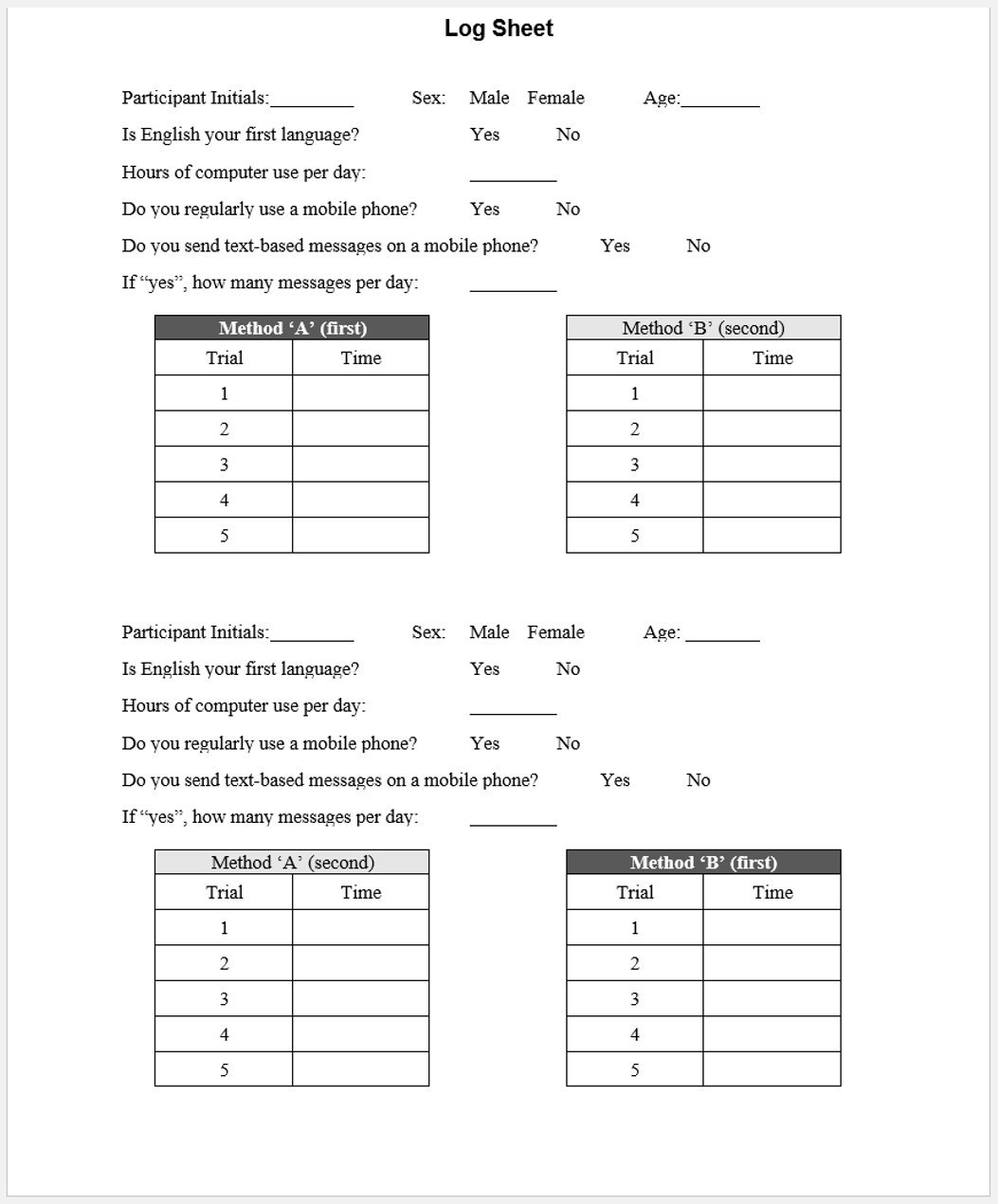}
\caption{Two-page handout for the in-class experiment.}
\Description{Two-page handout for the in-class experiment. The first page shows images of the Opti and Qwerty keyboard layouts used in the experiment.  Opt is labelled A and Qwerty is labelled B.  Below each layout of the phrase of text to enter: the quick brown fox jumps over the lazy dog.  The second page is for data collection including demographic data for age and gender.  There is a section for each participant.  For each keyboard layout there is a field to enter the time in seconds it took to enter the phrase.}
\label{fig:handout}
\end{figure}

Early in session 1, participants are divided into groups of two and participate in an experiment.  A hand-out is distributed for the in-class experiment.  See Fig.~\ref{fig:handout}.

\begin{figure}
\includegraphics[width=0.45\textwidth]{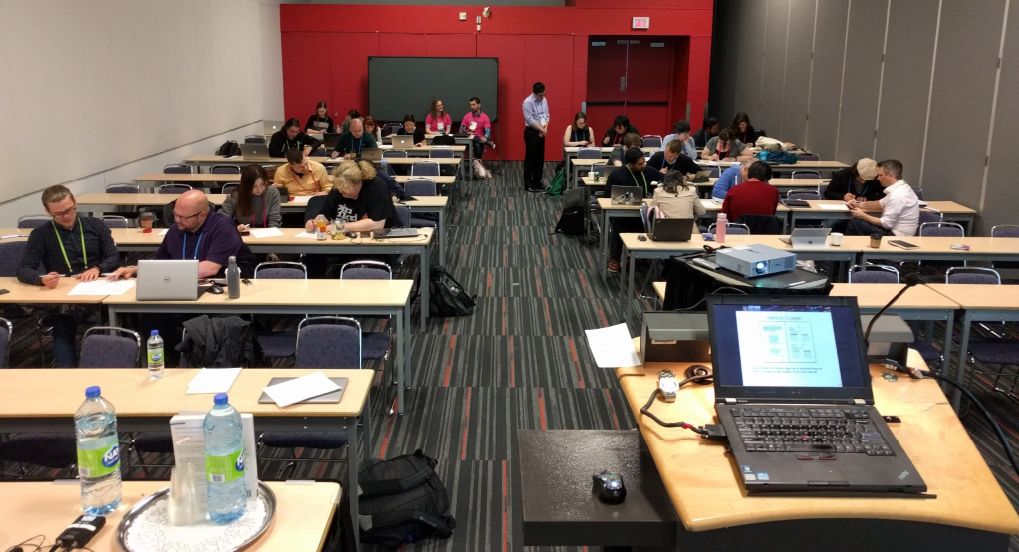}
\caption{In-class experiment for this course at a previous CHI conference.}
\Description{In-class experiment for this course at a previous CHI conference.  The photo shows a classroom with participants working in groups of two doing the in-class experiment.}
\label{fig:inclassexperiment}
\end{figure}

Following brief instructions, the in-class experiment proceeds.  During the experiment, participants take turns acting as a "participant" and as an "investigator". The participant does an experimental task -- entering a text phrase five times with a non-marking stylus on the image of a soft keyboard -- while the investigator measures the time to enter each phrase.  This is done twice, once for keyboard layout "A" and once for keyboard layout "B".  See Fig.~\ref{fig:inclassexperiment}. The data are entered in a log sheet.  When finished, the participant and investigator switch roles and the process is repeated.  This time the order of using the keyboard layouts is reversed, "B" first, then "A".  This is an example of \textit{counterbalancing}, as explained during the course.

As well as performance data, demographic information is entered on the log sheet.  The in-class experiment takes about 20 minutes.

Student volunteers (SVs) collect the hand-out sheets, leave the room, and transcribe the data from the handout sheets into a boilerplate spreadsheet, provided by the instructors. This is done as the course continues. Transcribing the data takes about 20-30 minutes with two SVs; i.e., one reads-out the data while the other inputs the data. This procedure has proved successful in previous offerings of this course. 

\begin{figure}
\includegraphics[width=0.3\textwidth]{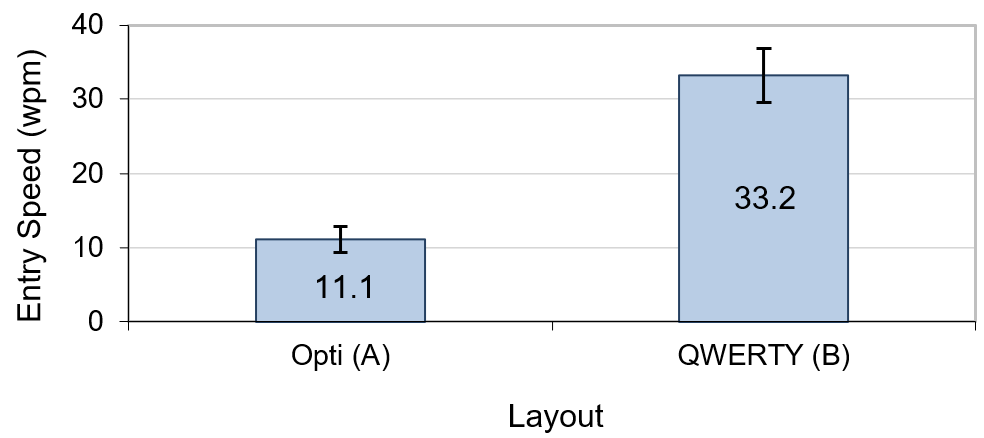} 
\includegraphics[width=0.3\textwidth]{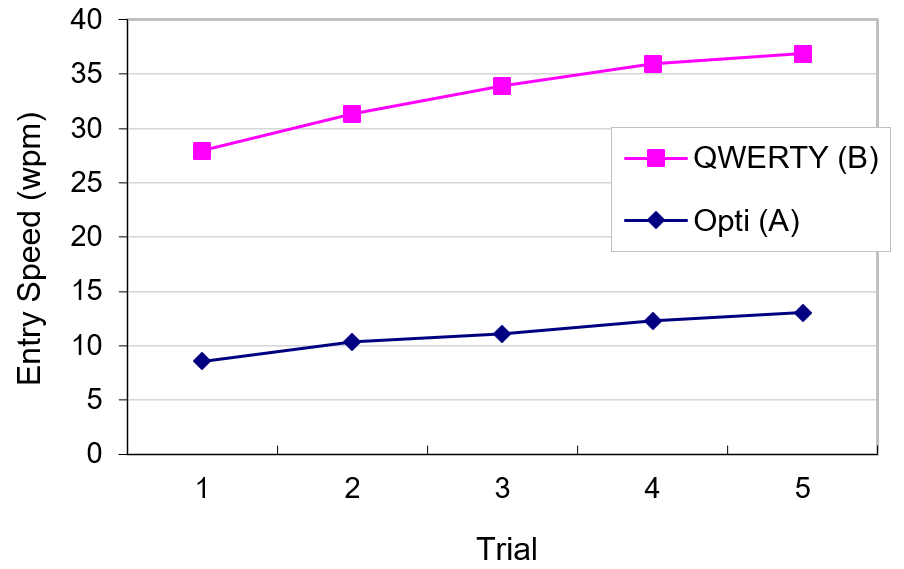}
\includegraphics[width=0.39\textwidth]{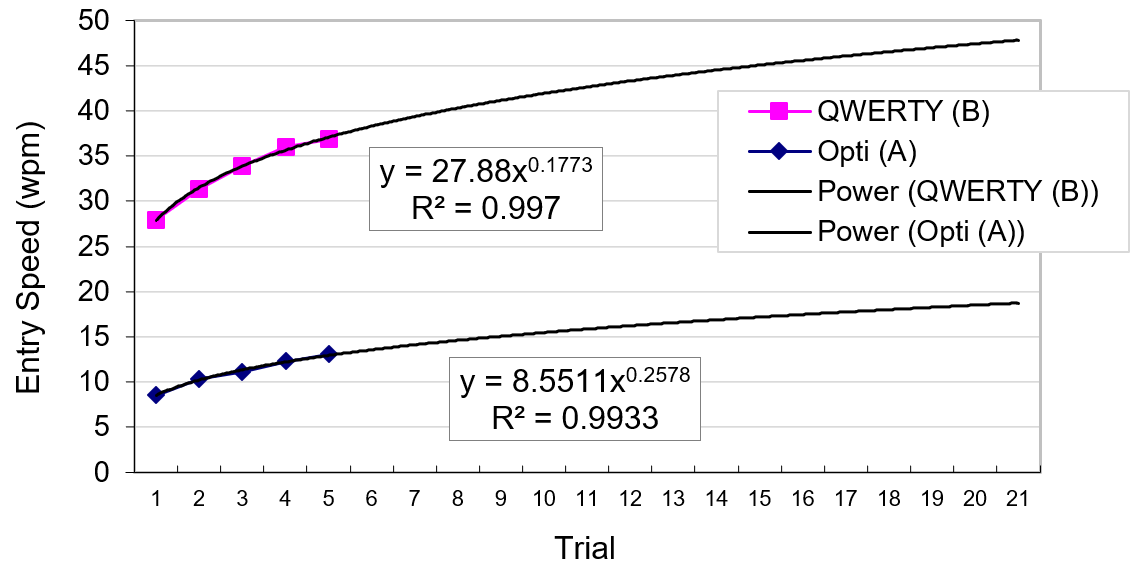}
\caption{Results from this course at a previous CHI conference. See text for discussion.}
\Description{Results from this course at a previous CHI conference. See text for discussion.  The figure contains three charts including a bar chart showing the entry speed for Opti versus Qwerty, a line chart showing the entry speed for Opti versus Qwerty over five trials, and a line chart showing the power law or learning for each keyboard layout and with an extrapolation to the 20 trials.}
\label{fig:results}
\end{figure}

During session 2, the course continues but now uses the methodology and results of the in-class user study to reinforce topics in the course.  Examples of the results are shown in Fig.~\ref{fig:results}.  The particular results are not important here.  However, it is extremely useful from a pedagogical perspective that the results discussed are from an experiment in which the course attendees have just participated.  Results of an analysis of variance are also presented.

\section{Instructor background}

Scott MacKenzie's research is in human-computer interaction with an emphasis on human performance, experimental methods and evaluation, interaction devices and techniques, etc.  He has more than 200 peer-reviewed publications in the field of Human-Computer Interaction (including more than 50 from the ACM's annual SIGCHI conference). In 2015, he was elected into the ACM SIGCHI Academy.  Full details: http://www.yorku.ca/mack/

Janet Read and Matt Horton have previously delivered courses at CHI on Child-Computer Interaction. For the last 15 years Janet has taught a course on research methods where she has used some of the aspects that are delivered in this tutorial and Matt has taught an advanced level course in user studies in HCI where he has expected students to plan experimental user studies.  Full details: https://chici.org/about/

\section{Resources}
Attendees needn't bring any resources.  Hand-outs will be disseminated during the course.

\section{Accessibility}
Attendees in need of accessibility arrangements are encouraged to contact the course organizers.  Appropriate assistance will be provided in consultation with the conference organizers.


\bibliographystyle{ACM-Reference-Format}
\bibliography{references}

@book{mackenzie2013a,
    author={MacKenzie, I. Scott},
    year={2013},
    title={Human-Computer Interaction: An Empirical Research Perspective},
    publisher={Morgan Kaufmann (an imprent of Elsevier)},
    address={Amsterdam}
}


\end{document}